\documentclass{aastex}
\usepackage{spr-astr-addons}

\RequirePackage{color}

\begin{document}

\title{A NLTE model atmosphere analysis of the pulsating sdO star SDSS
J1600+0748}


\author{M. Latour} \and \author{G. Fontaine} \and \author{P. Brassard}
\affil{D\'{e}partement de physique, Universit\'{e} de Montr\'{e}al, C.P. 6128, Succursale Centre-Ville, Montr\'{e}al, QC H3C 3J7, Canada}
\email{marilyn@astro.umontreal.ca}
\and
\author{P. Chayer}
\affil{Space Telescope Science Institute, 3700 San Martin Drive, Baltimore, MD 21218, USA}
\and 
\author{E. M. Green}
\affil{Department of Astronomy and Steward Observatory, University of Arizona, 933 North Cherry Avenue, Tucson, AZ 85721, USA}

\shorttitle{NTLE model atmospheres for sdO}
\shortauthors{Latour et al.}

\begin{abstract}
 We started a program to construct several grids of suitable model
 atmospheres and synthetic spectra for hot subdwarf O stars computed,
 for comparative purposes, in LTE, NLTE, with and without metals. For
 the moment, we use our grids to perform fits on our spectrum of SDSS
 J160043.6+074802.9 (J1600+0748 for short), this unique pulsating sdO
 star. Our best fit is currently obtained with NLTE model atmospheres
 including carbon, nitrogen and oxygen in solar abundances, which leads
 to the following  parameters for SDSS J1600+0748 : ${\it T}_{\rm eff}$
 = 69 060 $\pm$ 2080 K, log $g$ = 6.00 $\pm$ 0.09 and log {\it N}(He)/{\it
 N}(H) = $-$0.61 $\pm$ 0.06.  Improvements are needed, however,
 particularly for fitting the available He~\textsc{ii} lines. It is
 hoped that the inclusion of Fe will help remedy the situation. 
\end{abstract}


\section{Introduction}
SDSS J1600+0748 is this unique hot subdwarf O star showing short-period
p-mode instabilities (Woudt et al. 2006). Because this star is likely to
contain its fair share of metals (currently in unknown abundances and
proportions), we felt that it was important to assess the role of such
metals on its derived atmospheric parameters and pin down better its
position in the surface gravity-effective temperature plane. This is
particularly important for constraining an eventual seismic model (see
Fontaine et al. 2008). So, with this idea in mind, we have started the
construction of several grids of model atmospheres and synthetic spectra
suitable for the analysis of the optical spectrum that we gathered for
this star. 
\begin{figure}
\begin{center}
\includegraphics[scale=0.30,angle=270]{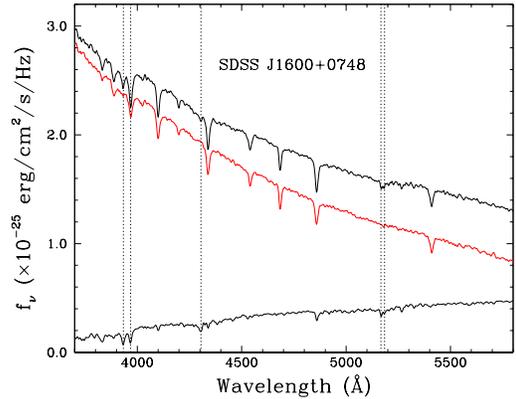}
\caption{Observed spectrum of the SDSS J1600+0748 system (top curve);
  template spectrum of the G0V companion (bottom curve); cleansed
  spectrum (middle curve, in red).}
\end{center}
\end{figure}

\section{Observational data}

Figure 1 shows our high-sensitivity ({\it S/N} $\thicksim$ 300),
low-resolution (8.7 \AA) optical spectrum of the SDSS J1600+0748 system
(upper curve) obtained at the Steward Observatory 2.3-m telescope over
four nights for a total exposure time of 12.23 h. Doppler shifts due to orbital or
proper motions of the star are corrected during the reduction procedure.
Note that this spectrum has also been presented in Fontaine et al. 2008. As first pointed out
by Woudt et al. (2006), and as can be seen here, this spectrum is
obviously contaminated by the light of a cool star showing the
Ca~\textsc{ii} K and H doublet at 3964-3968 \AA, the G band around 4305
\AA \ and the Mg~\textsc{i} complex at 5167-5184 \AA \ as its most
conspicuous features. One of us (E.M.G.) went to considerable efforts to
determine the most probable spectral type of that cool companion,
converging, at the end of the process, to a G0V star (see Fontaine et
al. 2008). A template G0V spectrum obtained with the same experimental
setup, and normalized properly in flux, is shown by the lower curve. The
resulting spectrum, obtained from the subtraction of those two, is shown
by the middle curve in red. It is this ``cleansed'' spectrum that forms
the observational basis of our study. It turns out, however, that the
atmospheric parameters derived from the analyses of the polluted
spectrum (upper curve) and the cleansed spectrum (middle curve) are not
hugely different.


\begin{figure}
\begin{center}
\includegraphics[scale=0.32,angle=270]{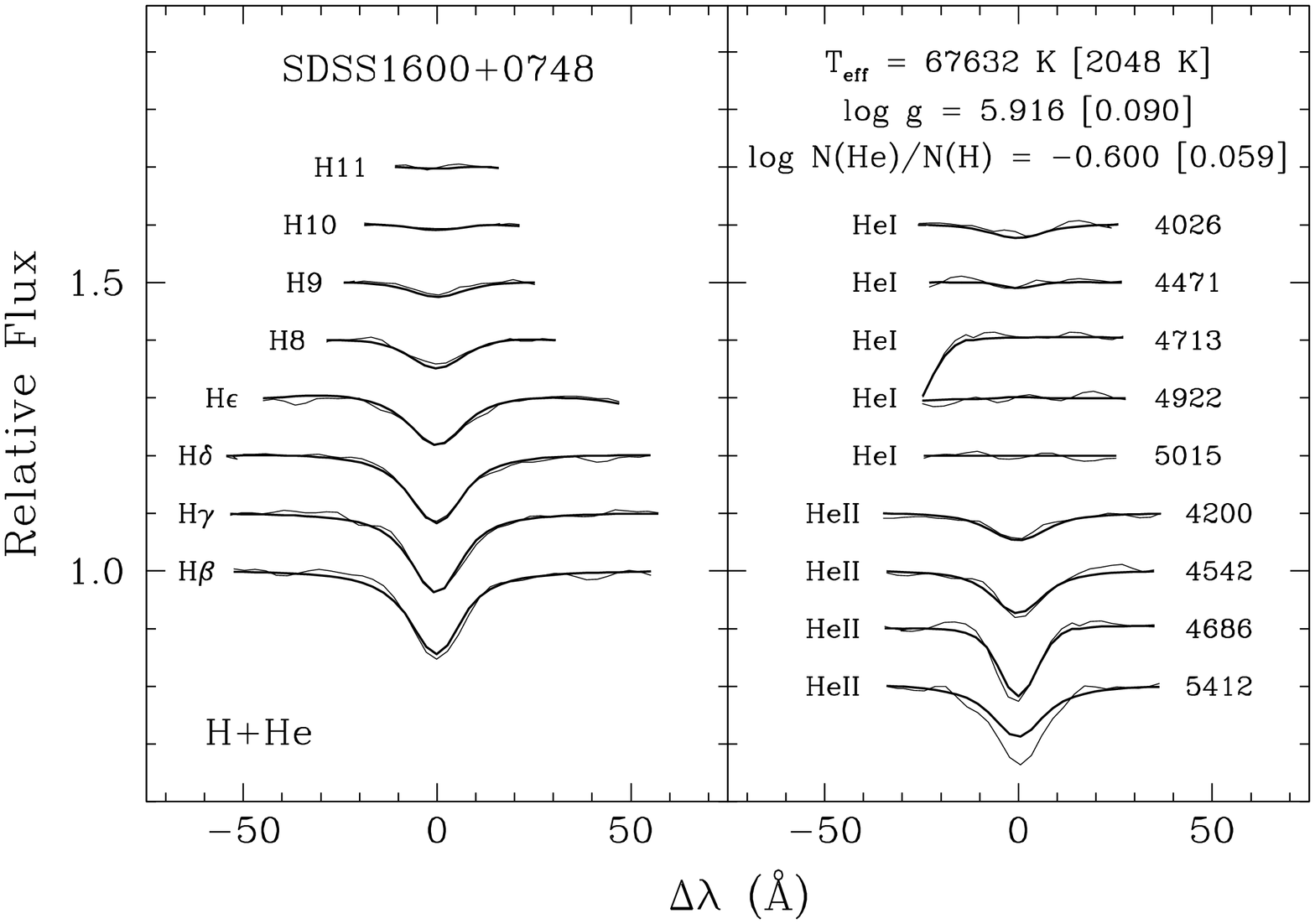}
\includegraphics[scale=0.32,angle=270]{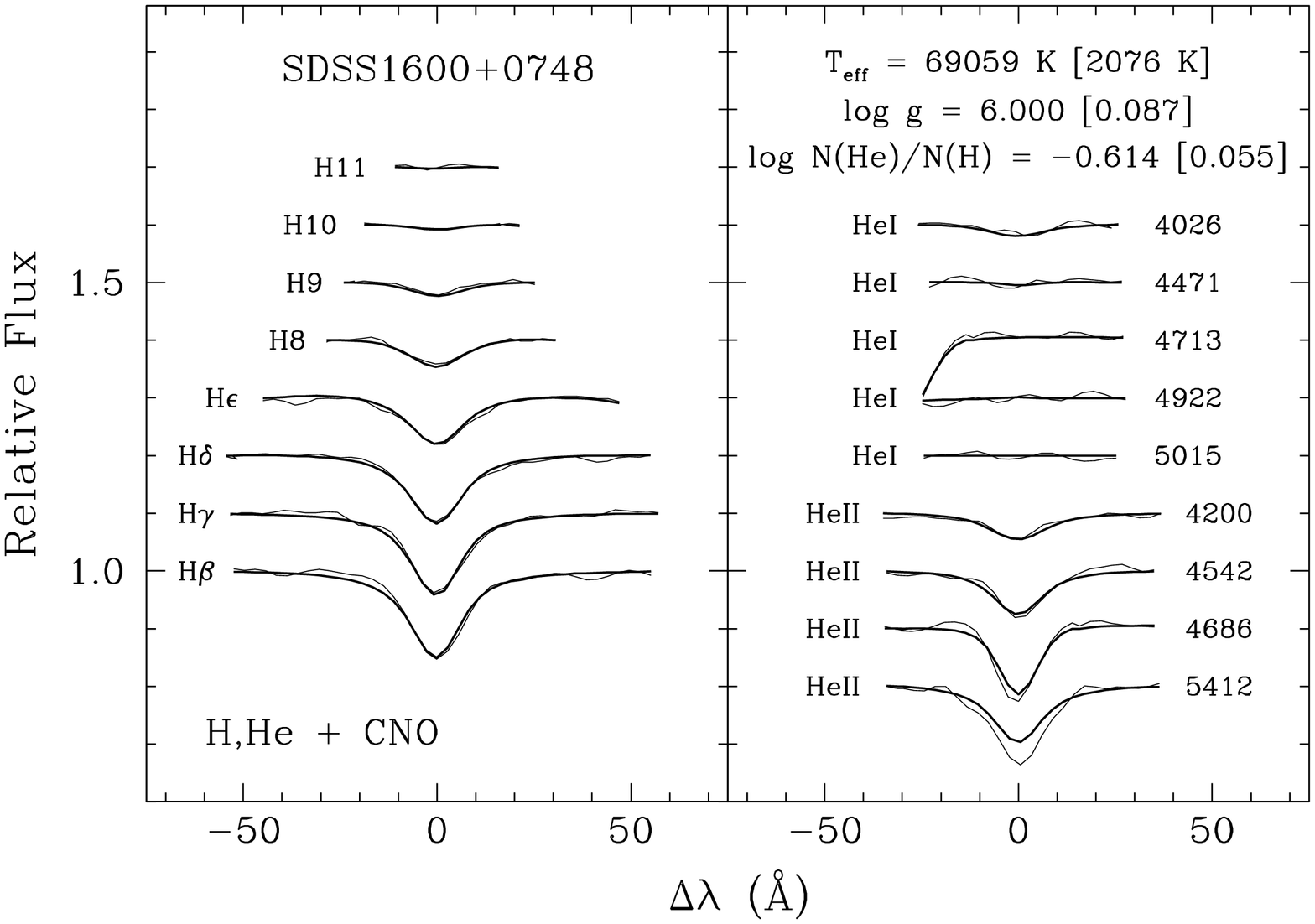}
\caption{Resulting fit of the spectrum of SDSS J1600+0748 with NLTE model
  atmospheres including H and He (top) and with CNO (bottom)} 
\end{center}
\end{figure}

\section{Grids of model atmospheres and synthetic spectra}

Only a few spectral analyses of hot subdwarf O stars including both
deviations from local thermodynamic equilibrium (NLTE) and metal line
blanketing have been done in the past (e.g., Lanz, Hubeny \& Heap 1997;
Deetjen 2000). This is essentially because calculating such model
atmospheres is very time consuming, especially when one includes
iron-peak elements. Thus, to compute our grids of NLTE model atmospheres
with metals, we have adapted the public codes TLUSTY and SYNSPEC of Ivan
Hubeny and Thierry Lanz to run in parallel on CALYS, our small cluster
of dedicated PC’s at the Universit\'{e} de Montr\'{e}al,
currently made up of 80 fast processors. This considerable effort was
carried out by one of us (P.B.).

At this time, we have two grids of NLTE models: one with H and He, and a
second one with these two elements as well as carbon, nitrogen and
oxygen (CNO for short) in solar abundances. In LTE, only the H and He
grid is now complete. Each grid is defined in terms of the effective
temperature (from 60 000 K to 80 000 K in steps of 2 000 K), the surface
gravity (log $g$ of 4.8 to 6.4 in steps of 0.2 dex), and the
helium-to-hydrogen number ratio (log {\it N}(He)/{\it N}(H) from $-$4.0
to 0.0 in steps of 0.5 dex). 
Our model atmospheres include model atoms available on TLUSTY web site \footnote{http://nova.astro.umd.edu/Tlusty2002/tlusty-frames-data.html}. 
We include the following 
ions in our models : C~\textsc{ii} to C~\textsc{v}, N~\textsc{ii} to N~\textsc{vi} and
O~\textsc{ii} to O~\textsc{vi}. For each ion there are a few models available with different number of levels and superlevels. In our grids, we took those used in the BSTAR2006 grid from Lanz \& Hubeny (2006).
Among the sdO stars, there's no ``standard'' chemical abundance unlike the trend observed in sdB stars (Blanchette et al. 2008). Specifically for SDSS J1600+0748 there's no hint, in currently available spectra, on which elements (heavier than helium) could be present in its atmosphere and even less on their abondance. Because of this lack of information, we used in our model a typical solar abundance for CNO. We are aware that these abundances are probably not the real ones and that further observations are necessary to get more information about the atomic species present in this star. Thus, we planned to get visible spectrum at 1 \AA \ resolution and a UV spectrum with the COS spectrograph on HST.

\begin{table*}
\begin{center}
\small
\caption{Results of our fitting procedure}
\begin{tabular}{@{}ccccc@{}}
\tableline
Grid & ${\it T}_{\rm eff}$ & log $g$ & log {\it N}(He)/{\it N}(H) & Spectrum \\ 
\tableline
LTE, H + He &51 056 $\pm$ 1 400 & 6.04 $\pm$ 0.08 & $-$0.71 $\pm$ 0.06 &cleansed\\
NLTE, H + He &67 632 $\pm$ 2 048 &5.92 $\pm$ 0.09 &$-$0.60 $\pm$ 0.06 &cleansed\\
NLTE, H + He + CNO &69 059 $\pm$ 2 076 &6.00 $\pm$ 0.09 &$-$0.61 $\pm$ 0.06 &cleansed\\
NLTE, H + He + CNO &70 681 $\pm$ 3 523 &5.87 $\pm$ 0.12 &$-$0.84 $\pm$ 0.07 &polluted \\
\tableline
\end{tabular}
\end{center}
\end{table*}

\begin{figure}
\begin{center}
\plotone{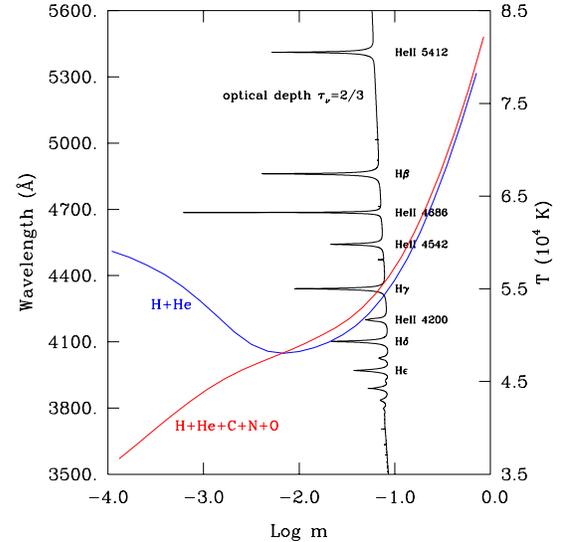}
\caption{Temperature and optical depth $\tau_{\nu}$ = 2/3 as functions of
  depth, where {\it m} is the column density, in the atmospheres of
  NLTE models with ${\it T}_{\rm eff}$ = 70 000 K, log $g$ = 6.0 and log
  {\it N}(He)/{\it N}(H) = $-$0.5.} 
\end{center}
\end{figure}

\section{Atmospheric parameters for SDSS J1600+0748}

We performed a simultaneous fit of the available Balmer lines and helium
lines in our observed spectrum using our different grids of synthetic
spectra. This was done using the standard line fitting method of
Bergeron et al. (1992). Results are summarized in Table 1, with errors
given on the parameters corresponding only to formal errors of the
fit. As mentioned before, we notice that the parameters obtained with
the polluted spectrum and the cleansed one are not that different. This is
reassuring. As for LTE model atmospheres, the quality of the fit is very
poor, leading to an effective temperature that is definitely too
small. Note that this result is quite uncertain as the temperature
predicted with such models is below the inferior limit of the grid. But
this experiment underlines the huge NLTE effects on the solution,
particularly on the effective temperature. On the other hand, inclusion
of CNO in solar abundances in NTLE models increases only slightly the
effective temperature and surface gravity obtained by the fitting 
procedure. These two results are illustrated in Figure 2. The fits to
H$\beta$ and He~\textsc{ii} (5412 \AA) lines are improved by the
inclusion of CNO, but the two strongest He~\textsc{ii} lines still
remain poorly fitted, showing a flux deficit compared to our best-fit
model.

\section{Effects of CNO line blanketing on our NLTE model atmospheres}

In stellar atmospheres, line blanketing of metal usually produce a drop
of temperature in the outermost layers (see, e.g., Bergeron, Saumon, \& Wesemael
1997). This effect is easily seen in Figure 3 where the red and blue
curves represent the temperature as a function of depth for model
atmospheres with and without CNO, respectively. We picked models in our
grids with parameters similar to those determined for J1600+0748. This
figure also features another interesting characteristic of our model
atmospheres with CNO, namely, the depth in the atmosphere where the
optical depth is equal to 2/3 as a function of the wavelength. With the
help of this curve, we can figure out where in the atmosphere are formed
the Balmer and He~\textsc{ii} lines as well as the temperature at which
these lines are formed.

\begin{figure}
\begin{center}
\plotone{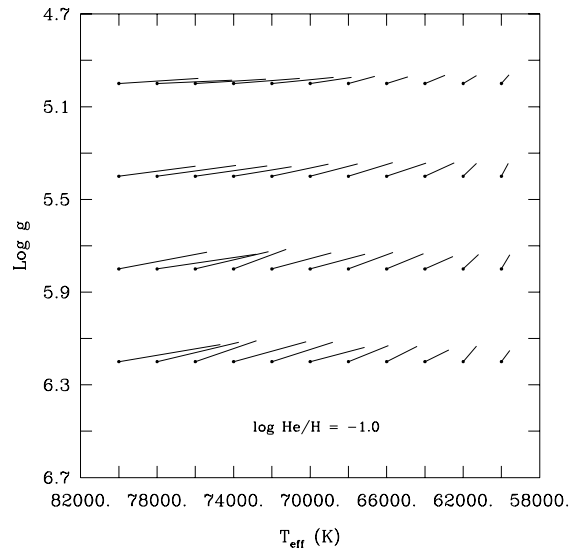}
\caption{ Results of the fitting procedure of our synthetic spectra with CNO using our grid of synthetic spectra with H and He only. See more details in the text.}

\end{center}
\end{figure}

We also carried out a detailed comparison between some of our synthetic
spectra characterized by a value of log {\it N}(He)/{\it N}(H) = $-$1.0
and a resolution of 8.7 \AA~ obtained by convolution to mimic
observational data. We considered our model spectra with CNO as
``observed'' spectra (represented by dots in Fig. 4) and performed a 
fit on each model with our grid of synthetic spectra computed with
models including H and He only. We used the same fitting technique as
for J1600+0748, but this time the value of log {\it N}(He)/{\it N}(H)
was fixed to $-$1.0, and searched for an optimal solution in terms of
temperature and gravity. The end of each line segment in Figure 4 gives
the result of the fitting procedure. We observe that the inclusion of
CNO in NLTE model atmospheres tends to increase the resulting effective
temperature and surface gravity, at least in our range of temperature,
with a larger effect at high ${\it T}_{\rm eff}$. Interestingly, our
results seem to point toward an inversed temperature trend at ${\it
  T}_{\rm eff}$ smaller than 60 000 K, a result that also depends on the
He abundance.  

\section{What is left to do}

The next planned stage in the construction of our grids of model atmospheres is
the inclusion of iron in NLTE models. We expect that a fit of J1600+0748
with this new grid will result in a (slightly?) larger effective
temperature as iron adds an important source of opacity. We hope being
able to improve the fit to the He~\textsc{ii} lines with the inclusion
of iron in our models. Since He~\textsc{ii} 5412 line is a transition
starting at the fourth energy level, while the three other strong HeII
lines start at the third level, an increase of the effective temperature
should raise the population of the fourth level and maybe cause a drop
of flux in our predicted He~\textsc{ii} 5412 line. Our ultimate goal is
to provide a convincing fit to the observed spectrum of SDSS J1600+0748
and, thus, reliable estimates of its atmospheric parameters. This is of
clear importance in the context of the future seismic exploration of the 
properties of SDSS J1600+0748, as spectroscopic constraints (especially
on the effective temperature) have proven to be essential during the
course of asteroseismological studies of pulsating hot subdwarf stars
(see, e.g., Charpinet et al. 2005).

\end{document}